\documentstyle[sprocl]{article}
\newcommand{\newc}{\newcommand}

\newc{\gsim}{\lower.7ex\hbox{$\;\stackrel{\textstyle>}{\sim}\;$}}
\newc{\lsim}{\lower.7ex\hbox{$\;\stackrel{\textstyle<}{\sim}\;$}}

\def\beq{\begin{equation}}
\def\eeq{\end{equation}}
\def\beqn{\begin{eqnarray}}
\def\eeqn{\end{eqnarray}}
\def\half{{\textstyle{1\over 2}}}
\def\calM{{\cal M}}

\def\inbar{\,\vrule height1.5ex width.4pt depth0pt}

\def\IC{\relax\hbox{$\inbar\kern-.3em{\rm C}$}}
\def\IQ{\relax\hbox{$\inbar\kern-.3em{\rm Q}$}}
\def\IR{\relax{\rm I\kern-.18em R}}
 \font\cmss=cmss10 \font\cmsss=cmss10 at 7pt
\def\IZ{\relax\ifmmode\mathchoice
 {\hbox{\cmss Z\kern-.4em Z}}{\hbox{\cmss Z\kern-.4em Z}}
 {\lower.9pt\hbox{\cmsss Z\kern-.4em Z}}
 {\lower1.2pt\hbox{\cmsss Z\kern-.4em Z}}\else{\cmss Z\kern-.4em Z}\fi}
\def\NPB#1#2#3{{\it Nucl.\ Phys.}\/ {\bf B#1} (19#2) #3}
\def\PLB#1#2#3{{\it Phys.\ Lett.}\/ {\bf B#1} (19#2) #3}
\def\PRD#1#2#3{{\it Phys.\ Rev.}\/ {\bf D#1} (19#2) #3}

\def\etal{{\it et al.\/}}

\def\beq{\begin{equation}}
\def\eeq{\end{equation}}
\def\beqn{\begin{eqnarray}}
\def\eeqn{\end{eqnarray}}
\def\ie{{\it i.e.}\/}
\def\eg{{\it e.g.}\/}

\begin{document}
\rightline{\tt hep-ph/0004129}
\rightline{April 2000} 
\bigskip
\title{NEW DIRECTIONS FOR NEW DIMENSIONS:\\
    FROM STRINGS TO NEUTRINOS TO AXIONS TO...~\footnote{
    Invited plenary talk given at {\it PASCOS '99:
       7}\/$^{\it th}$ {\it International Symposium on
       Particles, Strings, and Cosmology}\/
       (held at Lake Tahoe, California, 10--16 December 1999).
       To appear in the Proceedings.}}
\author{Keith R. Dienes}
\address{~\\
      Theory Division, CERN, CH-1211 Geneva 23, Switzerland\\
      Department of Physics, University of Arizona, Tucson, AZ  85721 
           USA~\footnote{Current address.  E-mail:  dienes@physics.arizona.edu}}
\maketitle\abstracts{
    In this talk, I discuss recent developments concerning
   the possibility of large extra spacetime dimensions.
   After briefly reviewing how such dimensions
   can lower the fundamental GUT, Planck, and string scales,
   I then outline how these scenarios lead to a
   new higher-dimensional seesaw mechanism for
   generating neutrino oscillations --- perhaps even 
   without neutrino masses.
   I also discuss how extra dimensions lead to
   new mechanisms contributing to the ``invisibility'' of the QCD axion.
   This talk reports on work done in collaboration with Emilian Dudas
     and Tony Gherghetta.
    } 
\input epsf.tex

\section{Introduction: Lowering the fundamental scales of physics}
\setcounter{footnote}{0}

The possibility of large extra spacetime dimensions has
recently received considerable
attention.  This is clearly an exciting prospect. 
One of the earliest proponents of TeV-scale 
extra dimensions was Antoniadis~\cite{Antoniadis},
who attempted to use such extra dimensions to explain
supersymmetry breaking.
Later, Witten~\cite{Witten} pointed out that extra
large dimensions could lower the string scale below
its usual value near 10$^{18}$ GeV, and
subsequently Lykken~\cite{Lykken} proposed that 
Witten's idea could be extended to lower the string scale
all the way to the TeV-range.
Finally, in March 1998, it was proposed that extra
dimensions could also be used to lower the fundamental
Planck scale~\cite{ADD} as well as the fundamental
GUT scale~\cite{DDG}.  Thus, combining these different
proposals, it becomes possible to contemplate a 
self-consistent scenario
in which all high fundamental energy scales
(GUT, Planck, and string scales)
are eliminated in favor of large extra spacetime dimensions!

It is important to distinguish two different types of
extra spacetime dimensions.  First, there are 
so-called ``universal'' extra dimensions.  These extra
dimensions are experienced by {\it all}\/ forces,
both gauge and gravitational;
in technical terminology, these 
extra dimensions are ``in the brane''.
Because they affect the gauge forces (as probed
by accelerator experiments),
such dimensions can be no larger than 
roughly an inverse TeV.
By contrast, the second class of extra dimensions
are felt only by gravity;
they are perpendicular to the D-brane on which the
gauge forces are localized, and may therefore be considered
``off the brane''.  
The sizes of such extra dimensions are significantly
less constrained, and may in fact
be as large as a millimeter.

Both of these types of extra dimensions play a role
in lowering the fundamental scales of physics.
Indeed, as outlined above, there are three different
proposals:  extra dimensions to lower the GUT scale~\cite{DDG},
extra dimensions to lower the Planck scale~\cite{ADD},
and extra dimensions to lower the string 
scale~\cite{Witten,Lykken,ADD,ShiuTye,DDG}.
We shall now briefly review these three proposals.

\begin{figure}[ht]
\centerline{ \epsfxsize 3.0 truein \epsfbox {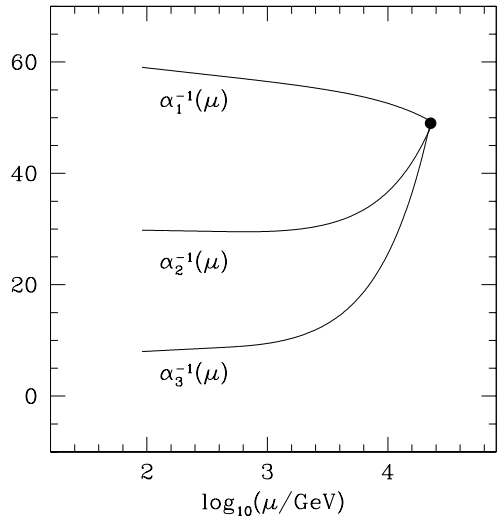}}
\caption{Unification of gauge couplings at
     the new unification scale $M'_{\rm GUT}\approx 20$ TeV,
    assuming the appearance of a single extra spacetime dimension
    of radius $R^{-1}=1$ TeV.}
\label{figureone}
\end{figure}

In the proposal of Ref.~\cite{DDG} to lower the GUT scale, 
one introduces some number
$\delta$ of ``universal'' extra spacetime dimensions ``in the brane'' 
[so that the Standard Model resides on a $D(3+\delta)$ brane],
and imagines that these dimensions have a common radius $R$.
Because these extra dimensions are felt by the gauge forces,
they change the running of the three gauge couplings from logarithmic
to power-law behavior:
\beq
     \alpha_i^{-1}(\mu) ~=~
      \alpha_i^{-1}(\mu_0)
      ~-~ {b_i-\tilde b_i \over 2\pi} \,\ln(R\mu)
           ~-~ {\tilde b_i X_\delta\over 2\pi \delta} \,\left\lbrack
                 \left(R\mu\right)^\delta -1\right\rbrack~.
\label{geffd}
\eeq
The emergence of power-law behavior is expected simply from
dimensional analysis, since the gauge couplings themselves
become dimensionful in higher dimensions, and hence have a classical
scaling in addition to their quantum-mechanical (logarithmic)
running.  This power-law behavior can also be realized
via a Kaluza-Klein summation, as discussed in Refs.~\cite{Taylor,DDG}. 
In Eq.~(\ref{geffd}), $X_\delta$ is a normalization
constant and ($b_i$, $\tilde b_i$) represent
the one-loop beta-functions appropriate for the zero-mode and excited
Kaluza-Klein states respectively.
The exact values of these beta-functions depend on details of
the compactification, as discussed in Ref.~\cite{DDG}.
However, as shown in Ref.~\cite{DDG}, the 
remarkable feature of this higher-dimensional running is 
that gauge coupling unification is still generally preserved, but with
a lowered unification scale!
As an interesting case, let us consider $R^{-1} = 1$ TeV
and $\delta=1$.  With one-loop beta-function coefficients
$(b_1,b_2,b_3)=(33/5, 1,-3)$ and $(\tilde b_1, \tilde b_2, \tilde b_3)=(3/5,-3,-6)$,
corresponding to a certain orbifold compactification 
discussed in Ref.~\cite{DDG},
we then find the unification shown
in Fig.~\ref{figureone}.
An important point to notice is that no large hierarchy is
needed between the scale of the extra dimensions and the
lowered GUT scale.

This reduced-scale unification leads to many important 
quantitative questions.  How predictive is this unification?
How perturbative is it?  How sensitive is it to unification-scale
effects?  What about higher-loop corrections?  These issues are
discussed in Ref.~\cite{DDG3}.
The upshot is that this sort of unification
scenario is predictive, perturbative, and 
not unreasonably sensitive to unification-scale effects.
There have also been many further extensions of these
basic ideas~\cite{others1}.
These include the study of two- and higher-loop effects;
the incorporation of extra matter beyond the MSSM
in order to increase the numerical accuracy of the unification;
alternative derivations of these RGE's from a Wilsonian perspective;
studies of regularization independence;
the extension of these ideas to the power-law running
of Yukawa couplings;
the higher-dimensional evolution of soft supersymmetry-breaking
mass parameters;
multi-step higher-dimensional unification scenarios;
and alternative embeddings of the Standard
Model into higher dimensions.
Alternative ideas pertaining to
reduced-scale gauge unification
have also been discussed in Refs.~\cite{others2}.

Extra dimensions can also be used to lower the Planck scale,
as pointed out in Ref.~\cite{ADD}.
Indeed, in many respects this Planck-scale proposal and
the above GUT-scale proposal are the gravitational/gauge
counterparts of each other.
Whereas the GUT proposal utilizes $\delta$ extra dimensions
``in the brane'' with radius $R$ to modify the running of the 
three {\it gauge}\/ couplings, 
the Planck proposal of Ref.~\cite{ADD} utilizes some number $n$ of
extra dimensions of radius $r$ ``off the brane'' to modify the running of the
effective dimensionless {\it gravitational}\/ (Newton) 
coupling $\tilde G_N(\mu)\equiv \mu^2 G_N$.  
As expected, the presence of the extra dimensions enhances
the power-law running of this gravitational coupling, 
changing the scaling behavior from $\mu^2$ to $\mu^{2+n}$.
This in turn lowers the Planck scale [\ie, the fundamental gravitational scale,
defined as the scale where 
$\tilde G_N(\mu)\sim {\cal O}(1)$].
Unlike the GUT proposal, however, one typically requires a 
significant hierarchy between the scale of the extra dimensions and 
the lowered Planck scale.  For example, in the case $n=2$ with 
a lowered Planck scale in the TeV-range,
one finds $r\approx$ millimeter $\approx (10^{-4}$ eV$)^{-1}$.
This hierarchy is exactly as large as the original hierarchy
between the electroweak scale and the usual four-dimensional
Planck scale.\footnote{
   As pointed out in Ref.~\cite{RS}, it may be possible 
   to avoid the former hierarchy and nevertheless explain
   the latter hierarchy by virtue of a ``warp'' rescaling factor.
   Issues surrounding gauge coupling unification in this scenario
   are discussed in Ref.~\cite{DDGanomaly}.}

\begin{figure}[ht]
\centerline{ \epsfxsize 3.0 truein \epsfbox {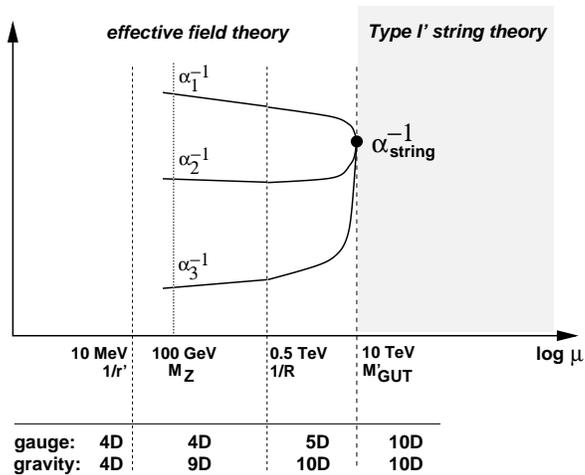}}
\caption{ The unification of gauge and gravitational couplings within
     the framework of a Type~I string theory, assuming a string
     scale at $M_{\rm string}=10$ TeV.  }
\label{bigembedding}
\end{figure}

Finally, combinations of both types of extra dimensions can
be used to lower the string scale~\cite{ShiuTye,DDG}.
For Type~I strings, the string scale 
ultimately depends on the six-volume $V_6$ of compactification
from ten flat dimensions to four flat dimensions:
\beq
          M_{\rm string} ~\sim~
         \sqrt{ 1\over \alpha_{\rm GUT} M_{\rm Planck}}
                        ~V_6^{-1/4}~.
\label{type1new}
\eeq 
Therefore, as discussed in Ref.~\cite{DDG},
if we seek to combine the above GUT and Planck  
scenarios together within string theory, we can write
$V_6=R^\delta r^{6-\delta}$ where $(R,\delta)$ describe
the extra dimensions ``in'' the brane (to produce a lowered
GUT scale) and $(r,6-\delta)$
describe the extra dimensions ``off'' the brane (to produce
a lowered Planck scale).
If we demand that the lowered GUT scale coincide with $M_{\rm string}$,
one can then solve to obtain a self-consistent solution.
For example, let us consider 
the case with $\delta=1$ and 
$R^{-1}\approx 0.5$ TeV.  (This extreme value may already be ruled out
experimentally, but will serve our illustrative purposes.)
This implies that
$ M'_{\rm GUT}\approx 10$ TeV,
which in turn implies (after a T-duality transformation) 
that the remaining five extra dimensions must have radius
     $r \approx (10~{\rm MeV})^{-1}$.

Thus, putting the pieces together in this example, we are led to a 
unified embedding into string theory, 
as illustrated in Fig.~\ref{bigembedding}.
Above the string scale $M_{\rm string}=10$ TeV,
the physics is described in terms of a full Type~I string theory.
Below 10 TeV, by contrast, the physics is described by a series
of  effective field theories in which the gauge and gravitational
forces feel different numbers of spacetime dimensions. 
Together, everything is balanced so as to produce a self-consistent
simultaneous lowering of GUT, Planck, and string scales.
Of course, other configurations are also possible.

\section{Light neutrino masses without heavy mass scales:\\
A higher-dimensional seesaw mechanism}
\setcounter{footnote}{0}

As we have seen,
the lesson from the above developments has been that
heavy mass scales in four dimensions can be replaced by
lighter mass scales in higher dimensions.
However, low-energy neutrino data seem to 
provide independent evidence for yet another heavy mass scale, 
namely the seesaw scale.  The seesaw mechanism relies on the 
existence of a new heavy mass scale $M\approx M_{\rm GUT}$ associated with a 
right-handed neutrino singlet field $N$.
The question then emerges whether it is possible to
generate light neutrino masses {\it without}\/ the introduction of 
a heavy mass scale, perhaps by some intrinsically higher-dimensional
mechanism.

To date, there have been essentially two 
ideas concerning how this might be accomplished
within the large extra-dimension framework:
one proposal~\cite{DDGneutrinos} utilizes a higher-dimensional seesaw mechanism,
and the other~\cite{ADDneutrinos} utilizes a higher-dimensional volume factor.
Both proposals originate with the same observation:
because the right-handed neutrino is a Standard-Model gauge singlet,
it need not be restricted to a ``brane'' with respect to the
full higher-dimensional space.
It is therefore possible for this field
to experience extra spacetime dimensions and thereby
accrue an infinite tower of Kaluza-Klein excitations.
This then leads to a number of higher-dimensional 
mechanisms~\cite{DDGneutrinos,ADDneutrinos,otherneutrinos}
for suppressing the resulting neutrino mass without a heavy mass scale.
In the following, we shall concentrate on one of the mechanisms advanced in
Ref.~\cite{DDGneutrinos}, namely the intriguing possibility that in higher
dimensions, neutrino oscillations need not imply the existence of neutrino
masses at all!  This would then eliminate the need for a high fundamental
scale.  Other mechanisms for explaining light but non-zero neutrino masses
are also discussed in Refs.~\cite{DDGneutrinos,ADDneutrinos,otherneutrinos}.
 
We begin by assuming that the right-handed neutrino feels
extra dimensions, while the left-handed neutrino $\nu_L$ does not.
For concreteness, we consider a  
Dirac fermion $\Psi$ in five dimensions, 
and work in the Weyl basis in which $\Psi$
can be decomposed
into two two-component spinors:  $\Psi = (\psi_1,\bar\psi_2)^T$.
When the extra spacetime dimension is compactified on
a $\IZ_2$ orbifold, it is natural for one of the two-component
Weyl spinors, \eg, $\psi_1$, to be taken to be even
under the $\IZ_2$ action $y\to -y$, while the other spinor
$\psi_2$ is taken to be odd.
If the left-handed neutrino $\nu_L$ is restricted
to a brane located at the orbifold fixed point $y=0$,
then $\psi_2$ vanishes at this point and so the most
natural coupling is between $\nu_L$ and $\psi_1$.
For generality, we will also include
a possible ``bare'' Majorana mass term for 
$\Psi$ of the form $\half M_0 \bar\Psi^{\rm c}\Psi$.
This then results in a Lagrangian of the form
\beqn
  {\cal L} &=& \int d^{4} x \,dy ~M_s\, \biggl\lbrace
     {\bar\psi}_1 i{\bar\sigma}^\mu \partial_\mu \psi_1
       +{\bar\psi}_2 i{\bar\sigma}^\mu \partial_\mu \psi_2
      + \half M_0 \left(
      \psi_1 \psi_1 + \psi_2 \psi_2 + {\rm h.c.} \right)\biggr\rbrace
                                 \nonumber\\
      && +~\int d^4 x ~ \biggl\lbrace
         {\bar\nu}_L i{\bar\sigma}^\mu D_\mu \nu_L
         ~+~({\hat m} \nu_L \psi_1|_{y=0} + {\rm h.c.})     \biggr\rbrace~
\label{klag}
\eeqn
where $y$ is the coordinate
of the extra compactified spacetime dimension
and where $M_s$ is the
mass scale of the higher-dimensional fundamental
theory (\eg, a reduced Type~I string scale).
Note that the last term represents
the Dirac brane/bulk Yukawa coupling between $\nu_L$ and $\psi_1$.
 
Next, we compactify the Lagrangian (\ref{klag}) down to four dimensions
by expanding the five-dimensional $\Psi$ field in Kaluza-Klein
modes.  Imposing the orbifold relations
$\psi_{1,2}(-y)=\pm \psi_{1,2}(y)$
implies that our Kaluza-Klein decomposition takes the form
     $\psi_1(x,y) = (2\pi R)^{-1/2}\sum_{n=0}^\infty
         \psi_1^{(n)}(x)\cos (ny/R)$
and a similar result for $\psi_2$ with cosine replaced by sine.
For convenience, we shall also define the linear combinations
$N^{(n)}\equiv(\psi_1^{(n)}+\psi_2^{(n)})/\sqrt{2}$
and
$M^{(n)}\equiv(\psi_1^{(n)}-\psi_2^{(n)})/\sqrt{2}$
for all $n>0$.
Inserting this decomposition into Eq.~(\ref{klag}) and
integrating over the compactified dimension, we then
obtain an effective four-dimensional Lagrangian 
in which the Standard-Model neutrino
$\nu_L$ mixes with the entire tower of Kaluza-Klein states
of the higher-dimensional $\Psi$ field with a mass mixing matrix
of the form
\beq
      \calM ~=~ \pmatrix{
         0 &  m   &   m  &   m &   m  &  m  &
         \ldots \cr
         m &  M_0 &   0  &   0  &   0  &  0  & \ldots \cr
         m  &  0   &   M_0+1/R  &   0  &   0  &  0  & \ldots \cr
       m &  0   &   0  &   M_0-1/R  &   0  &  0  & \ldots \cr
       m &  0   &   0  &   0  &   M_0+2/R  &  0  & \ldots \cr
       m &  0   &   0  &   0  &   0   &  M_0-2/R  & \ldots \cr
         \vdots  &  \vdots &   \vdots  &   \vdots
            &   \vdots &  \vdots  & \ddots \cr}~.
\label{newmatrix}
\eeq
In Eq.~(\ref{newmatrix}), we have defined the basis
        $(\nu_L, \psi_1^{(0)}, N^{(1)}, M^{(1)},
                  N^{(2)}, M^{(2)}, ...)$.
Note that $m\equiv \hat m/ \sqrt{2\pi R M_s}$ is the Dirac coupling
suppressed by a volume factor corresponding to the extra spacetime
dimension.

While in principle any value for $M_0$ is allowed (depending on the
structure of the full effective Lagrangian derived from the
particular underlying string model),
the topological constraints associated with 
Scherk-Schwarz compactification  
naturally suggest two specific values:   
$M_0=0$ (corresponding to no breaking of lepton number), and $M_0=(2R)^{-1}$
(corresponding to a global breaking of lepton number).
Let us here consider the non-trivial possibility $M_0=(2R)^{-1}$.
Note that this value is fixed topologically, and hence does not require
any fine-tuning.
It is then possible to solve for the eigenvalues and eigenstates
of the mass mixing matrix in Eq.~(\ref{newmatrix}).  Remarkably, 
it turns out that for any value of
$mR$, there exists an exactly zero eigenvalue, with a corresponding
mass eigenstate given exactly by 
\beq
       |\tilde \nu_L\rangle ~=~ {1\over \sqrt{ 1+ \pi^2 m^2 R^2}} \,
          \left\lbrace
         | \nu_L \rangle  ~-~  mR \,\sum_{k=1}^\infty
         {1\over k-1/2}\,\left\lbrack
            |N^{(k-1)}\rangle -|M^{(k)}\rangle
                   \right\rbrack \right\rbrace~
\label{eigenvec2}
\eeq
where we have defined $N^{(0)}\equiv \psi_1^{(0)}$.
Even though this result is exact for all $mR$, in most realistic
scenarios (see Ref.~\cite{DDGneutrinos}), we have $mR\ll 1$.
Thus, we see that 
even though this neutrino mass eigenstate contains
a small, non-trivial admixture of Kaluza-Klein states, 
the dominant component of our massless neutrino eigenstate remains the  
left-handed gauge-eigenstate neutrino $\nu_L$, as required phenomenologically.
Nevertheless, this particular admixture of excited Kaluza-Klein states
has rendered the neutrino eigenstate exactly massless!
In other words, the effects of the infinite tower of
Kaluza-Klein states for the $\Psi$ field have driven the neutrino mass
exactly to zero. 
 
\begin{figure}[ht]
      \centerline{\epsfxsize 3.0 truein \epsfbox {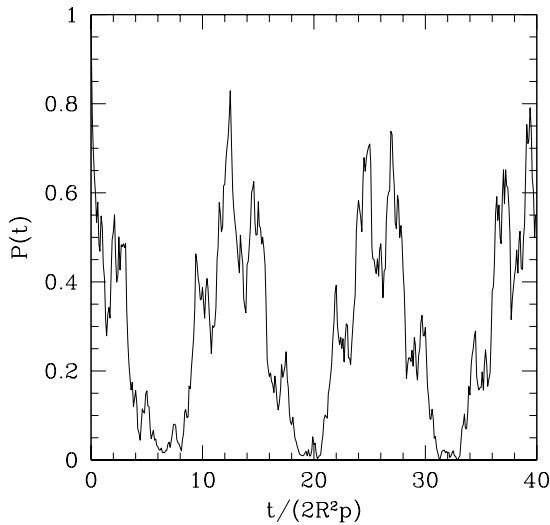}}
\caption{
     Higher-dimensional neutrino/anti-neutrino oscillations, 
      even when the neutrino itself is massless.
     Here we have plotted the probability that the gauge
     neutrino eigenstate $\nu_L$ is preserved as a function of time.
     The multi-component nature of
     the mixing between $\nu_L$ and the infinite set of 
      right-handed Kaluza-Klein neutrinos is reflected in
      the jagged shape of the oscillations.
       Note that the neutrino deficits are total
      even though the neutrino regenerations
       are only partial.  See Ref.~\protect\cite{DDGneutrinos} for further details.}
\label{oscfig2}
\end{figure}

One might still worry that a vanishing neutrino mass is unacceptable
because of the recent evidence for neutrino oscillations.
However, even though the neutrino mass is vanishing in this
scenario, {\it there continue to exist oscillations because of the
non-trivial mixings between the left-handed neutrino and the infinite
tower of Kaluza-Klein states}.
Specifically, upon diagonalizing the mass matrix (\ref{newmatrix}) 
and calculating
the resulting oscillation probabilities in the usual way, 
we find~\cite{DDGneutrinos} the result shown in Fig.~\ref{oscfig2}.
This figure thus provides explicit verification
that neutrino oscillations do indeed occur, even
though the physical neutrino is exactly massless.
Of course, we have been discussing only 
the simple case of neutrino/anti-neutrino oscillations.
However, this mechanism can easily be generalized to include
the case of flavor oscillations as well.

At first sight, it may seem strange 
that 
we are able to have neutrino
oscillations without neutrino masses.
However, the crucial point is that once the infinite towers
of Kaluza-Klein states are included,
the mixing mass matrix~(\ref{newmatrix}) with $M_0=(2R)^{-1}$
yields a zero eigenvalue {\it without becoming diagonal}\/.  
This is not possible
in the usual four-dimensional neutrino/anti-neutrino scenarios,
where the analogous mixing matrix takes the simple
form $\pmatrix{0 & m \cr m & M_0\cr}$.
Indeed, the masses of the right-handed Kaluza-Klein states themselves
are sufficient
to generate the desired oscillations indirectly, even though these
Kaluza-Klein states are in the ``bulk'' rather than on the brane. 
Thus, if such a scenario can be realized within the context
of a fully realistic string model, then the recent
observations of neutrino oscillations can be re-interpreted
not as providing evidence for neutrino masses, but
rather as providing evidence for extra spacetime
dimensions!

\section{Extra dimensions and ``invisible'' axions}
\setcounter{footnote}{0}

Many of the above ideas are completely general, and apply to
other bulk fields as well.  Towards this end,
let us now discuss how extra spacetime
dimensions may contribute to the invisibility of the QCD axion.
Like the graviton and right-handed neutrino, the QCD    
axion is also a Standard-Model singlet.  The QCD axion is therefore
free to propagate into the bulk.  

Can this be used to lower
the fundamental Peccei-Quinn (PQ)  symmetry-breaking scale?
This issue has been investigated in Refs.~\cite{CTY,DDGaxions}.
As explicitly shown in Ref.~\cite{DDGaxions} (and first proposed in Ref.~\cite{ADD}), 
it is indeed possible to exploit the volume factor of large extra dimensions
in order to realize a large effective four-dimensional PQ scale
from a smaller, higher-dimensional fundamental PQ scale.  
Thus, once again, no large fundamental energy scales are required.

However, as discussed in Ref.~\cite{DDGaxions},
the presence of Kaluza-Klein axions can have important and
unexpected effects on axion phenomenology.
Just as in the neutrino case discussed above, the na\"\i ve
four-dimensional axion mixes with the infinite tower of Kaluza-Klein
axions, with a mass matrix given in Ref.~\cite{DDGaxions}.
This mixing has a number of interesting phenomenological consequences.
First, as shown in Ref.~\cite{DDGaxions}, under certain circumstances
the mass of the axion essentially {\it decouples}\/ from the PQ scale,
and instead is set by the radius of the extra spacetime dimension!
Thus, axions in the $10^{-4}$ eV mass range are consistent with
(sub-)millimeter extra dimensions.
This decoupling implies that it may be possible to adjust the
mass of the axion independently of its couplings to matter.
This is not possible in four dimensions.

\begin{figure}[ht]
\centerline{
      \epsfxsize  2.75 truein \epsfbox {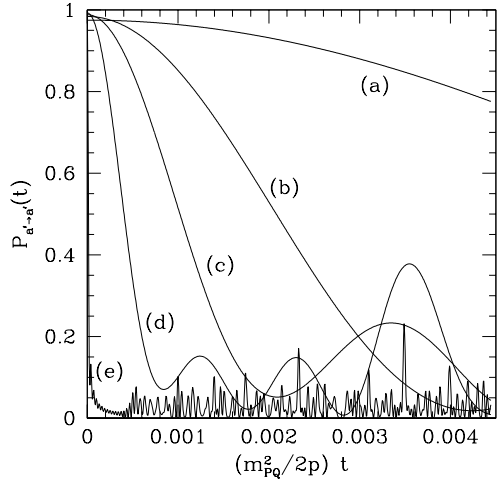}}
\caption{The axion preservation probability $P_{a'\to a'}(t)$
    as a function of the number $n_{\rm max}$ of axion Kaluza-Klein
    states which are included in the analysis.
    In this plot we show the results for
    (a) $n_{\rm max}=1$;   (b) $n_{\rm max}=2$;
         (c) $n_{\rm max}=3$;   (d) $n_{\rm max}=5$;  and   (e) $n_{\rm max}=30$.
    As $n_{\rm max}$ increases, the axion probability
    rapidly falls to zero as a result of the destructive
   interference of the Kaluza-Klein states,
   and remains suppressed
     without significant axion regeneration at any later times.
       This ``decoherence'' of the axion implies that there is
    negligible probability for subsequently detecting
     the original axion state at any future time.  Further
     details can be found in Ref.~\protect\cite{DDGaxions}.}
\label{dropping}
\end{figure}

Second, as discussed in Ref.~\cite{DDGaxions},
the usual four-dimensional axion is no longer a mass eigenstate  
because of the non-trivial axion mass mixing matrix. 
This implies that the four-dimensional axion should
undergo {\it laboratory  oscillations} which are entirely  
analogous to neutrino oscillations.  Moreover, because the axion is now 
a bulk field, Standard-Model particles couple not only
to the axion zero-mode, but rather to the entire linear superposition 
$a'\sim \sum_n a_n$ (where $a_n$ are the axion Kaluza-Klein modes).
Therefore, the quantity of phenomenological interest 
is the probability $P_{a'\to a'}(t)$ that $a'$ is preserved as a function of time.
This probability~\cite{DDGaxions} is shown in Fig.~\ref{dropping}.
Unlike the neutrino case, we see that the probability
drops rapidly from $1$ (at the initial time $t=0$) to extremely small
values (expected to be  $\approx 10^{-16}$ when an appropriately truncated
set of $10^{16}$ Kaluza-Klein
states are included in $a'$).  At no future time does this probability
regenerate.  Essentially, the axion state $a'$ has ``decohered''
and becomes invisible with respect to subsequent laboratory interactions.
This decoherence is therefore a possible mechanism contributing to an
invisible axion.

Finally, one can investigate the effects of Kaluza-Klein axions
on cosmological relic axion oscillations.  In this regard it is important
to understand whether the coupled axion Kaluza-Klein states accelerate or retard
the dissipation of the cosmological energy density associated with
these oscillations.  Remarkably, one finds~\cite{DDGaxions}
that the net effect of these coupled Kaluza-Klein axions is to 
either {\it preserve}\/ or {\it enhance}\/ the rate of energy dissipation. 
This implies that the usual relic oscillation bounds are loosened
in higher dimensions,
which suggests that it may be possible to raise the 
effective PQ symmetry-breaking scale beyond its usual four-dimensional value.  
This could therefore potentially
serve as another factor contributing to axion invisibility. 
Together, these results suggest that
it may be possible to develop a new, higher-dimensional approach
to axion phenomenology.

\section{Conclusions}
\setcounter{footnote}{0}

Only experiment will decide if large extra spacetime dimensions
actually exist, and if the fundamental high-energy scales of
physics are really as low as the TeV-range.  Nevertheless, 
what is remarkable about the recent developments is that they illustrate
that the fundamental energy scales are not immutable, and 
that the parameter space for physics 
beyond the Standard Model is significantly broader than had 
been previously thought. 
Moreover, it is equally remarkable and gratifying 
that ideas originally
born in string theory are having such a profound effect
on the answers to primarily phenomenological questions, 
and that these ideas may be potentially testable in 
the not-too-distant future.  
If nothing else, these may be the most valuable lessons
that we may take with us from the brane world.


\section*{Acknowledgments}

I wish to thank the organizers of the PASCOS '99 conference,
and especially Jack Gunion, for the opportunity to speak
at such a stimulating and wide-ranging conference.


\end{document}